\documentclass[reprint,aps,prl,showpacs]{revtex4-1}
\usepackage{natbib}
\usepackage{amssymb,amsmath,graphicx}
\begin{document}

\title{Adiabatic passage in the presence of noise}
\author{T. Noel}
\email{noelt@uw.edu}
\homepage{http://depts.washington.edu/qcomp/}
\author{M. R. Dietrich}
\altaffiliation{Current address: Physics Division, Argonne National Laboratory, Argonne, IL 60439}
\author{N. Kurz}
\altaffiliation{Current address: Nion Co., 1102 8th St Kirkland, WA 98033}
\author{G. Shu}
\altaffiliation{Current address: School of Chemistry, Georgia Institute of Technology, Atlanta, Georgia 30332}
\author{J. Wright}
\author{B. B. Blinov}
\affiliation{Department of Physics, University of Washington, Seattle, WA 98195}
\newcommand{\Ba}{\textsuperscript{138}Ba\textsuperscript{+}}

\begin{abstract}
We report on an experimental investigation of rapid adiabatic passage (RAP) in a trapped barium ion system. RAP is implemented on the transition from the $6S_{1/2}$ ground state to the metastable $5D_{5/2}$ level by applying a laser at 1.76~$\mu$m. We focus on the interplay of laser frequency noise and laser power in shaping the effectiveness of RAP, which is commonly assumed to be a robust tool for high efficiency population transfer. However, we note that reaching high state transfer fidelity requires a combination of small laser linewidth and large Rabi frequency.
\end{abstract}

\maketitle

\paragraph*{}  Adiabatic passage has been used for coherent population transfer in many fields \cite{bloch46,loy74,grischkowsky76,caussyn94}.  Nuclear magnetic resonance (NMR) was the first system in which adiabatic passage was implemented \cite{bloch46}.  In an NMR system, slowly varying the magnetic field frequency or direction can transfer population between nuclear spin states.  The technique has since been applied to infrared and optical transitions in atomic and molecular systems \cite{loy74,grischkowsky76}.  In such systems the frequency of a laser is swept across a transition, effecting population transfer between the ground state and an excited state.  More recently, adiabatic passage has been applied in trapped singly ionized alkali earth element systems similar to ours.  This has enabled high fidelity ($\geq 0.99$) readout of ionic qubit states \cite{wunderlich07, poschinger09}.  Using adiabatic passage for state detection has several important advantages over a simple $\pi$-pulse of resonant light.  First, small drifts in laser frequency do not cause a dramatic change in transfer fidelity.  Similarly, adiabatic passage is mostly insensitive to frequency shifts caused by magnetic field fluctuations.  Additionally, adiabatic passage efficiency is not strongly dependent on laser power, and so obviates the need for laser intensity stabilization.   On the other hand, adiabatic  passage is necessarily slower than a $\pi$-pulse and still requires optical coherence of a sufficient degree that Rabi oscillations can be observed.  Previous experiments on RAP were done with sufficiently high Rabi frequency and low driving field noise that the role of noise processes was not investigated.  There is considerable interest in the role that noise plays in adiabatic processes for applications in adiabatic quantum computing \cite{childs01,roland05,pekola10}.  Here we present a detailed study of the dependence of adiabatic passage efficiency on laser noise and other relevant parameters.

\paragraph*{}  The theory behind adiabatic passage has been described in several papers and books \cite{vitanov01, shore90,wunderlich07,jha10,cohen-tannoudji92}, so we will only sketch the idea in broad strokes.  Consider a Hamiltonian $H$, which describes two states whose energies cross as a function of a parameter $\Delta$.  Now add to this Hamiltonian a term which couples the two states, yielding a new Hamiltonian $H'$.  The eigenstates of $H'$ will exhibit an avoided crossing as a function of  $\Delta$.  Consider such a coupled system initially in an eigenstate of $H$ with $\Delta$ set such that the energy splitting between the states is much larger than the coupling energy.  If $\Delta$ is now varied sufficiently slowly across the energy level crossing, the system will adiabatically follow the eigenstates of $H'$ and population will be transferred to the other eigenstate of $H$.  This population transfer is referred to as rapid adiabatic passage.

\begin{figure}[ht]
\includegraphics[trim=80mm 0mm 0mm 0mm, width=60mm]{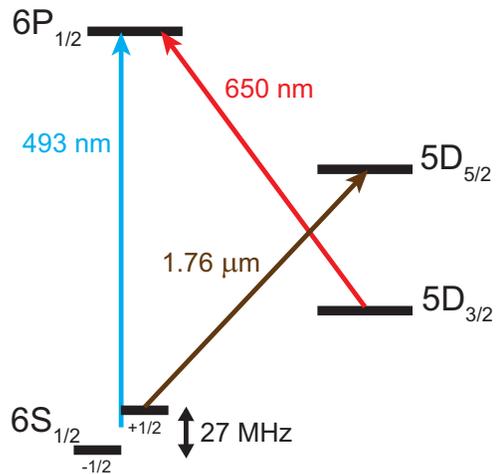}
\caption{Relevant energy levels of \Ba \ (not to scale) with magnetic sublevels suppressed in all levels other than the ground state.  (Color online.)}
\label{levels}
\end{figure}

\paragraph*{} Fig. \ref{levels} shows the energy level diagram for \Ba , which has no nuclear spin and so is devoid of hyperfine structure.  In our experiments the ion is trapped in a linear Paul trap with an axial secular frequency of 600~kHz and radial frequencies of 1.4~MHz and 2.6~MHz \cite{dietrich10}.  The ion is Doppler cooled with about \mbox{$3 \ \mu $W} of 493~nm light.  This light is derived from a frequency-doubled 986~nm external cavity diode laser (ECDL).  Since the branching ratio for decay into the long-lived \mbox{($\tau = 80$ s)} $5D_{3/2}$ state is about 0.25 we also repump the ion with about \mbox{$20 \ \mu$W} of 650~nm light provided by a second ECDL.  By sending in an additional 493 nm beam of the appropriate circular polarization along the axis of the magnetic field seen by the ion we are able to optically pump the ion into either Zeeman level of the ground state.  For a more detailed description of our apparatus see Ref. \cite{dietrich10}.  

\paragraph*{}
As shown in Fig. \ref{levels}, the energy difference between the $5D_{5/2}$ state and the ground state corresponds to \mbox{$1.76 \ \mu$m} light which is provided by a fiber laser (Koheras Adjustik$^{\mathrm{TM}}$).  The fiber laser output is split and half the power is sent through an I/Q modulated double-passed acousto-optic modulator (DPAOM) as part of a Pound-Drever-Hall cavity locking setup.  The other half of the power is sent through a single-passed AOM (SPAOM) and then to the ion.  The frequency shifts applied by the two AOM's allow us to finely tune the fiber laser frequency.  Additionally, by sending a chirped radio-frequency (RF) signal to the SPAOM we are able to write a frequency chirp onto the laser.  We initially used a Stanford Research Systems DS345 frequency synthesizer to produce the RF chirp, but the shortest chirp it could produce was 1~ms in duration.  Because this was too slow to probe the full relevant parameter space we built a field programmable gate array (FPGA) controlled direct digital synthesizer (DDS).

\paragraph*{} For investigation of adiabatic passage, our two-level system is provided by the $6S_{1/2},m_{J}=+1/2$ ground state and one of the Zeeman states of the $5D_{5/2}$ level.  The experimental sequence is as follows: (1) optically pump the ion into a specific ground state Zeeman level, (2) perform adiabatic passage to transfer the ion to the metastable $5D_{5/2}$ level, (3) read out the state of the ion.  State readout employs the fact that the $5D_{5/2}$ level is long-lived \mbox{($\tau = 35$ s)} and disjoint from the cooling cycle.  Thus if we address the ion with the cooling lasers we will quickly be able to distinguish between the dark $5D_{5/2}$ level and the bright ground state \cite{dietrich10}.  With standard refractive optics outside the vacuum chamber we can collect thousands of fluorescence photons per second from an ion in the cooling cycle, whereas the signal from laser scatter and photomultiplier tube (PMT) dark counts is less than one hundred counts per second.  We are therefore able to reliably distinguish a dark ion from a bright one in tens of milliseconds, long before the $5D_{5/2}$ level decays.

\begin{figure}[ht]
\includegraphics[width=85mm]{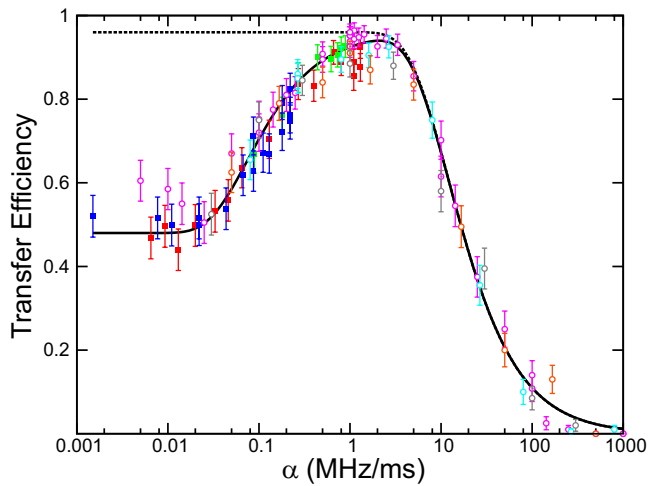}
\caption{RAP transfer efficiency plotted as a function of the frequency sweep rate.  The solid square data points were taken with frequency sweeps provided by a DS345 Stanford Research Systems function generator.  The open circle data points were taken using our homebuilt FPGA controlled DDS.  The colors of the various data points correspond to different data sets, which were taken with a variety of sweep widths between 0.2~MHz and 1.2~MHz.  The solid curve is a fit to the Lacour \emph{et al.} model \cite{Lacour07}, while the dashed line is a fit to the data with \mbox{$\alpha > 2$} using Landau-Zener theory \cite{zener32}.  The errorbars are statistical.  (Color online.)}
\label{effvsalpha}
\end{figure}

\paragraph*{} Using the setup described above, we were able to investigate the probability of successful adiabatic passage as a function of the 1.76~$\mu$m laser frequency sweep rate $\alpha$ and incident laser power.  Fig. \ref{effvsalpha} shows the probability of adiabatic passage as a function of $\alpha$ for various sweep widths.  Note that the data supports the prediction of the Landau-Zener model of adiabatic passage that the efficiency should only depend on the sweep rate, and not on the sweep width or sweep duration independently \cite{zener32}.
The solid curve is a fit based on the two level dephasing model of Lacour \emph{et al.} \cite{Lacour07}, in which the environment is assumed to act as a Markovian noise bath with small correlation times.  This leads to a master equation for the evolution of the two-state system, which includes the effect of the environment (entering our system primarily through laser frequency noise).  The Lacour \emph{et al.} model yields the following expression for the probability of adiabatic passage (after unit conversion and interpretation of parameters to our system): 

\begin{equation} 
P = F\left(\frac{1}{2}(1-e^{- 2 \pi^{2} \Gamma \Omega / \alpha}) + e^{-2 \pi^{2} \Gamma \Omega / \alpha}P_{LZ}\right)
\label{Lacour}
\end{equation}

In this expression, $P_{LZ}=1-exp(-\pi^{2} \Omega^{2} / \alpha)$ is the probability of adiabatic passage in the Landau-Zener model, $\Gamma$ is our 1.76~$\mu$m laser linewidth, $\Omega$ is the resonant Rabi frequency, $F$ is the optical pumping efficiency, and $\alpha$ is the sweep rate.  The fit shown in Fig. \ref{effvsalpha} uses $\Omega$, $\Gamma$, and $F$ as free parameters, and yields values of \mbox{$\Omega = 35$ kHz}, \mbox{$\Gamma = 110$ Hz} and $F=0.96$.  The fit has a $\chi^{2}$ per degree of freedom of 1.5.  Note the drastic reduction in the transfer efficiency below \mbox{$\alpha = 1$ MHz/ms}, where the Landau-Zener model would predict nearly perfect population transfer.  For the Rabi frequency and laser linewidth in our system, there appears to be an optimal operating range of alpha between approximately 0.5 and 3~MHz/ms.  Since the noise in our system starts to be visible on sweeps of about 1~ms in duration, the extracted linewidth includes noise frequencies above 1~kHz.  This procedure allows us to put a better upper bound on the value for our laser linewidth than we have found possible by interrogating the ion with a constant laser frequency.  The fitted value for optical pumping efficiency is a product of several factors: (1) Fig. \ref{effvsalpha} is a combination of many datasets, and we did not tune up our system as well as possible before taking each set, (2) the quarter waveplate used to create the required polarization for optical pumping is optimized for 488~nm rather than 493~nm where we use it, and (3) the windows on the vacuum chamber which houses the trap introduce small polarization errors due to stress-induced birefringence.

\begin{figure}[ht]
\includegraphics[width=85mm]{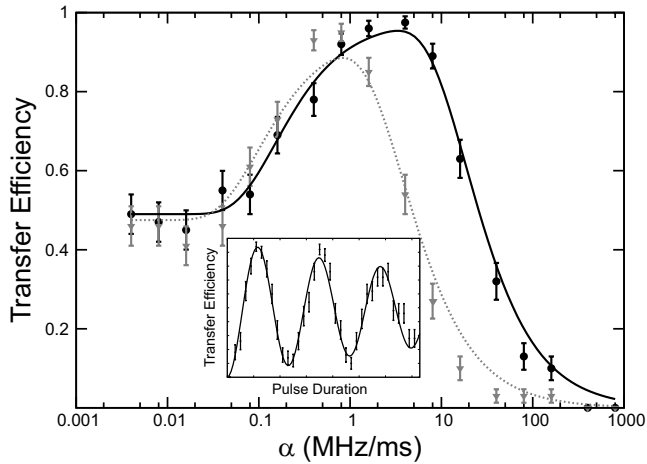}
\caption{RAP efficiency as a function of sweep rate for different Rabi frequencies.  The black data points correspond to full laser power (2.5~mW), which results in a Rabi frequency of 43~kHz.  The gray data points were taken with a Rabi frequency of 19~kHz, roughly a factor of four reduction in laser intensity.  The curves are again fits to the Lacour model, this time with only the laser linewidth and optical pumping efficiency as free parameters.  The black fit yields $\Gamma = 180$~Hz and $F = 0.98$ and the gray fit yields $\Gamma = 260$~Hz and $F = 0.95$.  The inset figure shows an example of Rabi  oscillation in our system, from which we can extract the Rabi frequency.  The decay in contrast is consistent with the Doppler-cooled ion temperature.  All errorbars are statistical.}
\label{appower}
\end{figure}

\begin{figure}[ht]
\includegraphics[width=85mm]{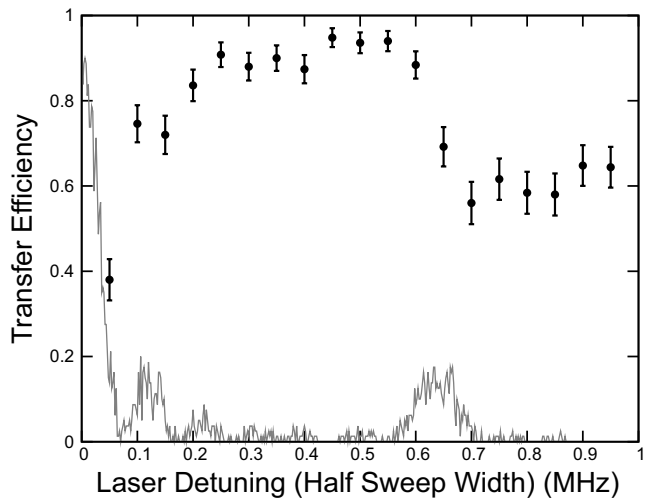}
\caption{RAP efficiency for different sweep widths, constant sweep duration of 0.9 ms.  The gray line shows the probability of driving the $6S_{1/2} \rightarrow 5D_{5/2}$ transition by constant frequency excitation.  The peak substructure is due to coherent excitation.  The sideband at $\sim$600~kHz is due to the harmonic motion of the ion in our trap.  The black data points show the transfer efficiency of RAP sweeps which are symmetric about the central peak.  The points are plotted at the frequency corresponding to half the RAP sweep width.  (A point plotted at 0.5~MHz corresponds to a RAP sweep starting at a detuning of -0.5~MHz and ending at 0.5~MHz.)  Note that when the sweep starts too close to resonance or when the sweep range includes the trap sidebands the transfer efficiency is low.  The errorbars are statistical.}
\label{effwidth}
\end{figure}

\paragraph*{}  We also study the role of laser power in determining the efficiency of RAP population transfer.  Some results of this study are shown in Fig. \ref{appower}.  The fits shown are again to the Lacour model [Eq. (\ref{Lacour})].  From data sets taken at various laser powers we obtain a range of fitting parameters for the laser linewidth  of $100$ Hz $\leq \Gamma \leq 300$ Hz.  This implies an uncertainty in the extracted linewidth which is larger than any of the uncertainties dictated by any individual fit.  This is likely due, in part, to drifts in the performance of the laser locking system.  Note the minimal dependence of RAP transfer efficiency on laser power near $\alpha = 1$~MHz/ms.  That the plateau of maximal transfer efficiency is largely independent of driving field intensity makes RAP robust against intensity noise.

\paragraph*{}  The presence of additional spectral features within the RAP frequency sweep can strongly affect the efficiency of population transfer.  In the special case of trapped ion systems, this means that the width of the RAP frequency sweep is limited by the presence of the secular sidebands.  Fig. \ref{effwidth} depicts this phenomenon.  The gray line shows the probability of transfer to the $5D_{5/2}$ state by constant frequency excitation for a resonant $\pi$-time.  The line separated from the central peak by just over 600 kHz is the axial secular frequency sideband.  Once the RAP frequency sweep extends beyond the sidebands the efficiency of state transfer is greatly reduced.  This clearly demonstrates that for successful RAP one must have a system with sufficiently separated spectral features.

\paragraph*{} These experimental results confirm the importance of driving field noise in determining the efficiency of RAP.  For applications of RAP that require high-fidelity, a large ratio of Rabi frequency to driving field noise is required.  Using the Lacour model, we propose the following as a simple rule of thumb: to acheive a transfer efficiency of 0.99 a ratio of $\Omega / \Gamma \gtrsim 10^{3}$ is necessary.  If a system meets this requirement and the reduction in speed involved in using an adiabatic technique is acceptable, then RAP is a good option for high-fidelity, robust population transfer.

\paragraph*{}  By systematically investigating the relevant parameter space, we have shown that RAP requires the ratio of dephasing rate to Rabi frequency to be small in order to use RAP to transfer population with high probability.  Additionally, we find the model of dephasing in two-state adiabatic passage due to Lacour \emph{et al.} to describe our results well.  By accepting fits to this model quantitatively, we are able to use the dependence of RAP on sweep rate to measure our 1.76~$\mu$m fiber laser linewidth to be \mbox{$\approx 200$ Hz}.

\paragraph*{} 
We would like to acknowledge the contributions to this work of Aaron Avril, Shaw-Pin Chen, Chen-Kuan Chou, Richard Graham, Matt Hoffman, Adam Kleczewski, and Paul Pham.  
This research was supported by the National Science Foundation Grants No. 0758025 and No. 0904004.

\end{document}